# Implementing WHERE and ORDER BY as spreadsheet formulas

Paul Mireault
SSMI International
Montréal, Canada
Paul.Mireault@SSMI.International

*Abstract*— **The WHERE and ORDER BY clauses of the SQL SELECT statement select a subset of rows in the result of a database query and present the result in the specified order. In a spreadsheet program like Microsoft Excel, one could use the filter and sort buttons, or use its Query or its Pivot Table tools to achieve a similar effect. The disadvantage of using those tools is that they don't react automatically to changes in the calculated values of the spreadsheet. In this paper, we develop spreadsheet formulas that implement SQL's WHERE and ORDER BY clauses.**

## I. Introduction

The similarity between spreadsheet models and relational databases has interested researchers for some time. Mireault [1] developed a spreadsheet development methodology based on the conceptual model of Information Systems.

Cunha et al. proposed a model-driven approach [2], and Cunha et al. proposed a method to reverse-engineer a spreadsheet to extract a relational model and proposed a method to do the opposite: generate a spreadsheet from a relational model [3]. Mireault, in [4] and [5], showed that a multi-dimensional spreadsheet has a structure similar to a data warehouse.

Finally, Mireault [6] presented a way to build spreadsheet formulas to perform aggregate calculations like the GROUP BY clause of a SELECT statement.

In this paper, we present spreadsheet formulas that function like the WHERE and ORDER BY clauses of a SELECT statement.

## II. Curent tools and their limitations

While Microsoft Excel has tools to manipulate data using sorts, filters, Pivot Tables and Queries, they require the use of macros or VBA programs or they need the user to perform some manipulations and, thus, they are not suitable for calculated models.

Google sheet has a query function that automatically updates its results when the underlying data changes. While this is an improvement over Excel, it is also limited to datasets that are organized vertically. Also, using this feature cannot be exported to an Excel spreadsheet.

When analyzing data, the Pivot Table itself is the result sought by the user. But in the case of a model, the user can perform different scenario analysis to explore the impact of some input variable values on some key result variables. If we were to use Pivot Tables to aggregate some variables, we would need one Pivot Table for each initial and final dimension sets. The result of one Pivot Table would be variables that are used in the calculation of other variables, which are in turn used in another Pivot Table.

But Pivot Tables and Queries do not update dynamically when the underlying values change. After modifying an input value, the spreadsheet user would then have to update each Pivot Table, in the proper order, to observe the correct result. While it is possible to program routines in VBA that would be triggered automatically when an input value changes, this requires training that most Excel users don't have.

Similarly, Excel's built-in tools for sorting and filtering only work on datasets presented vertically and are not dynamic: they require that the user perform the action when the values are changed.

In this paper, we will present worksheet structures and Excel formulas that will dynamically perform a selection and a sort like the WHERE and ORDER BY clauses of SQL's SELECT statement. The formulas can work with datasets presented vertically or horizontally.

## III. Example problem

Our example is based on a spreadsheet developed for a federation of sports clubs. The Federation acts as a centralized IT department for its clubs. It develops or purchases systems and manages them for the benefit of its clubs. For example, it has a Human Resources system, a Payroll system, a membership management system, a web portal system, etc. Clubs are usually too small and don't have the resources to have their own IT specialists and they rely on the Federation to handle all the technical details. The Federation sells the different systems, called *products*, to its affiliated clubs using a pricing formula based on the clubs' number of members or number of employees, called *population*. Some products, like HR and Payroll, use the employee population and others, like Membership and Portal, use the members population.

Clubs are free to buy the products they want. There are 72 clubs, 22 products and 1297 purchases. The club populations range from 3 to 10000 members.

## IV. IMPLEMENTING THE WHERE CLAUSE

The spreadsheet calculates the prices of the products bought by the clubs, based on input parameters that the Sales Director enters for each product, like the Floor Price and the Ceiling Price. Figure 1 shows the result for some particular input values. It has 1297 rows.

To extract rows of interest from the results, we first need to create an indicator variable that evaluates to 1 when the row needs to be selected and 0 otherwise. This is shown in Figure 2.

The second step is to create a sequence number that gets incremented when the indicator variable is 1. This is illustrated in Figure 3.

Finally, we prepare the extraction area by building a column with sequential numbers, like column E in Figure 4. We then identify, with the MATCH function the rows that are selected in the source worksheet:

```
=MATCH(E2,'Club-Product Results'!J:J,0)
```

The match type parameter 0 indicates that we want an *exact match*. In this case, the function returns the row where it first encounters the search value. Having evaluated the row number of each selected row in the source, we can extract all the desired column with a simple INDEX function, as shown in Figure 5.

Figure 6 illustrates the dynamic nature of the extraction by simply changing the value of the *Desired Product* cell to CV.

| | A | B | C | D | E | F | G | H |
|---|---|---|---|---|---|---|---|---|
| 1 | Club-Product Results | | Club ID | Product Code | Previous Cost | Prod-Club | Population Product-Club | Product Cost |
| 2 | | | | | | | | |
| 3 | | | 689 | PURCH | $1 527,00 | PURCH-689 | 422 | $2 000,00 |
| 4 | | | 689 | BIM | $10 374,00 | BIM-689 | 9 | $20 000,00 |
| 5 | | | 689 | DOFIN | $19 967,69 | DOFIN-689 | 422 | $20 000,00 |
| 6 | | | 689 | GPI | $13 467,85 | GPI-689 | 422 | $15 000,00 |
| 7 | | | 689 | JADE | $11 404,41 | JADE-689 | 422 | $12 000,00 |
| 8 | | | 689 | MEM | $5 821,19 | MEM-689 | 422 | $6 000,00 |
| 9 | | | 689 | MZK-AX | $13 002,00 | MZK-AX-689 | 422 | $15 000,00 |
| 10 | | | 689 | REP | $4 094,00 | REP-689 | 422 | $3 000,00 |
| 11 | | | 689 | SER | $8 354,00 | SER-689 | 422 | $5 000,00 |
| 12 | | | 689 | VIDEO | $335,00 | VIDEO-689 | 422 | $400,00 |

Figure 1 Final results of the spreadsheet model

RATE  fx  =IF(Product_Code=Desired_Product,1,0)

| | A | B | C | D | E | F | G | H | I | J | K | L |
|---|---|---|---|---|---|---|---|---|---|---|---|---|
| 1 | Club-Product Results | | Club ID | Product Code | Previous Cost | Prod-Club | Population Product-Club | Product Cost | Selected Product Indicator | Seq No of Selected Product | | |
| 2 | | | | | | | | | | 0 | | |
| 3 | | | | 689 | PURCH | $1,527.00 | PURCH-689 | 422 | $2,000.00 | =IF(Product_Code=Desired_Product,1,0) | 0 | |
| 4 | | | | 689 | BIM | $10,374.00 | BIM-689 | 9 | $20,000.00 | 0 | 0 | |
| 5 | | | | 689 | DOFIN | $19,967.69 | DOFIN-689 | 422 | $20,000.00 | 0 | 0 | |
| 6 | | | | 689 | GPI | $13,467.85 | GPI-689 | 422 | $15,000.00 | 1 | 1 | |
| 7 | | | | 689 | JADE | $11,404.41 | JADE-689 | 422 | $12,000.00 | 0 | 1 | |
| 8 | | | | 689 | MEM | $5,821.19 | MEM-689 | 422 | $6,000.00 | 0 | 1 | |
| 9 | | | | 689 | MZK-AX | $13,002.00 | MZK-AX-689 | 422 | $15,000.00 | 0 | 1 | |
| 10 | | | | 689 | REP | $4,094.00 | REP-689 | 422 | $3,000.00 | 0 | 1 | |
| 11 | | | | 689 | SER | $8,354.00 | SER-689 | 422 | $5,000.00 | 0 | 1 | |
| 12 | | | | 689 | VIDEO | $335.00 | VIDEO-689 | 422 | $400.00 | 0 | 1 | |
| 13 | | | | 711 | PURCH | $12,026.00 | PURCH-711 | 4128 | $6,638.07 | 0 | 1 | |
| 14 | | | | 711 | AG | $3,070.00 | AG-711 | 10 | $500.00 | 0 | 1 | |
| 15 | | | | 711 | BIM | $28,279.07 | BIM-711 | 623 | $20,000.00 | 0 | 1 | |
| 16 | | | | 711 | CONS | $3,976.00 | CONS-711 | 4128 | $4,232.00 | 0 | 1 | |
| 17 | | | | 711 | DOFIN | $46,243.54 | DOFIN-711 | 4128 | $31,520.00 | 0 | 1 | |
| 18 | | | | 711 | GEO | $8,403.00 | GEO-711 | 4128 | $7,754.67 | 0 | 1 | |
| 19 | | | | 711 | GPI | $21,912.18 | GPI-711 | 4128 | $25,143.16 | 1 | 2 | |
| 20 | | | | 711 | H | $3,639.82 | H-711 | 98 | $4,000.00 | 0 | 2 | |

Figure 2 Creation of Indicator variable

| | A | B | C | D | E | F | G | H | I | J |
|---|---|---|---|---|---|---|---|---|---|---|
| | | | | | | | | | | Seq No |
| | Club- | | | | | | Population | | Selected | of |
| | Product | | | Product | Previous | | Product- | Product | Product | Selected |
| 1 | Results | | Club ID | Code | Cost | Prod-Club | Club | Cost | Indicator | Product |
| 2 | | | | | | | | | | 0 |
| 3 | | | 689 | PURCH | $1,527.00 | PURCH-689 | 422 | $2,000.00 | 0 | =J2+I3 |
| 4 | | | 689 | BIM | $10,374.00 | BIM-689 | 9 | $20,000.00 | 0 | 0 |
| 5 | | | 689 | DOFIN | $19,967.69 | DOFIN-689 | 422 | $20,000.00 | 0 | 0 |
| 6 | | | 689 | GPI | $13,467.85 | GPI-689 | 422 | $15,000.00 | 1 | 1 |
| 7 | | | 689 | JADE | $11,404.41 | JADE-689 | 422 | $12,000.00 | 0 | 1 |
| 8 | | | 689 | MEM | $5,821.19 | MEM-689 | 422 | $6,000.00 | 0 | 1 |
| 9 | | | 689 | MZK-AX | $13,002.00 | MZK-AX-689 | 422 | $15,000.00 | 0 | 1 |
| 10 | | | 689 | REP | $4,094.00 | REP-689 | 422 | $3,000.00 | 0 | 1 |
| 11 | | | 689 | SER | $8,354.00 | SER-689 | 422 | $5,000.00 | 0 | 1 |
| 12 | | | 689 | VIDEO | $335.00 | VIDEO-689 | 422 | $400.00 | 0 | 1 |
| 13 | | | 711 | PURCH | $12,026.00 | PURCH-711 | 4128 | $6,638.07 | 0 | 1 |
| 14 | | | 711 | AG | $3,070.00 | AG-711 | 10 | $500.00 | 0 | 1 |
| 15 | | | 711 | BIM | $28,279.07 | BIM-711 | 623 | $20,000.00 | 0 | 1 |
| 16 | | | 711 | CONS | $3,976.00 | CONS-711 | 4128 | $4,232.00 | 0 | 1 |
| 17 | | | 711 | DOFIN | $46,243.54 | DOFIN-711 | 4128 | $31,520.00 | 0 | 1 |
| 18 | | | 711 | GEO | $8,403.00 | GEO-711 | 4128 | $7,754.67 | 0 | 1 |
| 19 | | | 711 | GPI | $21,912.18 | GPI-711 | 4128 | $25,143.16 | 1 | 2 |
| 20 | | | 711 | H | $3,639.82 | H-711 | 98 | $4,000.00 | 0 | 2 |

*Figure 3 Creation of the selection sequence number*

| | A | B | C | D | E | F | G | H | I | J |
|---|---|---|---|---|---|---|---|---|---|---|
| 1 | Product Cost by Club | | | | NoSeq | Row Num | Club | Pop | Cost | Previous Cost |
| 2 | | | | | 1 | =MATCH(E2,'Club-Product Results'!J:J,0) | | | | $13,467.85 |
| 3 | Desired Product | | GPI | | 2 | 19 | 711 | 4128 | $25,143.16 | $21,912.18 |
| 4 | | | | | 3 | 37 | 712 | 8328 | $41,353.68 | $43,562.10 |
| 5 | Optimal Combination of Desired Prod | | 8 | | 4 | 55 | 713 | 3137 | $21,318.25 | $17,758.00 |
| 6 | Combi-Prod | | 8-GPI | | 5 | 73 | 714 | 6098 | $32,746.67 | $32,557.72 |
| 7 | | | | | 6 | 93 | 721 | 6688 | $35,023.86 | $34,983.59 |
| 8 | Floor Price | | $15,000 | | 7 | 114 | 722 | 6398 | $33,904.56 | $35,667.66 |
| 9 | Ceiling Price | | $125,000 | | 8 | 133 | 723 | 10626 | $50,223.16 | $63,810.19 |
| 10 | Floor Population | | 1500 | | 9 | 151 | 724 | 7434 | $37,903.16 | $44,252.16 |
| 11 | Ceiling Population | | 30000 | | 10 | 170 | 731 | 2957 | $20,623.51 | $19,102.00 |
| 12 | Unit price between Floor and Ceiling | | $3.8596 | | 11 | 187 | 732 | 23500 | $99,912.28 | $83,563.15 |
| 13 | | | | | 12 | 207 | 733 | 11574 | $53,882.11 | $66,282.77 |
| 14 | | | | | 13 | 226 | 734 | 24585 | $104,100.00 | $88,718.40 |
| 15 | | | | | 14 | 244 | 735 | 5735 | $31,345.61 | $32,199.64 |
| 16 | | | | | 15 | 263 | 741 | 16007 | $70,991.93 | $72,988.54 |
| 17 | | | | | 16 | 283 | 742 | 8654 | $42,611.93 | $47,914.60 |
| 18 | | | | | 17 | 304 | 751 | 5978 | $32,283.51 | $31,269.72 |
| 19 | | | | | 18 | 324 | 752 | 18654 | $81,208.42 | $80,642.41 |
| 20 | | | | | 19 | 346 | 753 | 7868 | $39,578.25 | $43,562.76 |

*Figure 4 Extraction area*

| | E | F | G | H |
|---|---|---|---|---|
| 1 | NoSeq | Row Number | Club | Pop |
| 2 | 1 | =MATCH(E2,'Club-Product Results'!J:J,0) | =INDEX('Club-Product Results'!C:C,F2) | =INDEX('Club-Product Results'!G:G,F2) |
| 3 | 2 | =MATCH(E3,'Club-Product Results'!J:J,0) | =INDEX('Club-Product Results'!C:C,F3) | =INDEX('Club-Product Results'!G:G,F3) |
| 4 | 3 | =MATCH(E4,'Club-Product Results'!J:J,0) | =INDEX('Club-Product Results'!C:C,F4) | =INDEX('Club-Product Results'!G:G,F4) |

*Figure 5 Extraction of the selected row's values*

*Figure 6 Extraction for a different desired product*

## V. IMPLEMENTING THE ORDER BY CLAUSE

While being able to dynamically extract the correct rows as the spreadsheet is recalculated is useful by itself, a developer may want to present the results in a particular order. In a database environment, a developer will use a SELECT statement with an ORDER BY clause indicating the order in which the rows of the result are to be presented.

In the following sections, we will first consider the case with a simple ORDER BY with one column with unique values and then we will study the case where we have we have non-unique values and need to use more than one column to break the ties.

### A. ORDER BY with a single column of unique values

When we use the RANK.EQ function with unique values, it will return a set of values from 1 to the size of the set we are using; when there are equalities, it then assigns the same rank to each row with the same value. For example, if there are three rows tied for the first rank, the function will assign the value 1 to each of the three rows and 4 to the next one. Figure 7 illustrates the ranks calculated with the population column. The last parameter, with value 0, indicates a ranking from highest (1) to lowest (72 in this case.)

Finally, we use the same principles used in Figure 4 to obtain the results in the desired order, as shown in Figure 8.

### B. ORDER BY with non-unique values and multiple columns

When we do the same thing with the Cost column we run into the problem shown in Figure 9. Our data has 8 rows with the top cost of $125,000 and they have been assigned rank 1, so the MATCH function in column U cannot find ranks 2 to 8.

*Figure 7 Ordering according to the population*

## Figure 8 Ordering by decreasing population

Formula bar: `=MATCH(M2,Pop_Rank,0)` (cell shown as RATE/N2)

| Seq No | Row Number | Club | Pop | Cost | Previous Cost | Pop Rank | | Seq No Pop | Row Num Pop | Club | Pop | Cost |
|---|---|---|---|---|---|---|---|---|---|---|---|---|
| 1 | 6 | 689 | 422 | $15,000.00 | $13,467.85 | 72 | | 1 | =MATCH(M2,Pop_Rank,0) | | | |
| 2 | 19 | 711 | 4128 | $25,143.16 | $21,912.18 | 57 | | 2 | 23 | 763 | 46784 | $125,000.00 |
| 3 | 37 | 712 | 8328 | $41,353.68 | $43,562.10 | 33 | | 3 | 45 | 831 | 41406 | $125,000.00 |
| 4 | 55 | 713 | 3137 | $21,318.25 | $17,758.00 | 62 | | 4 | 48 | 851 | 35984 | $125,000.00 |
| 5 | 73 | 714 | 6098 | $32,746.67 | $32,557.72 | 42 | | 5 | 46 | 841 | 34901 | $125,000.00 |
| 6 | 93 | 721 | 6688 | $35,023.86 | $34,983.59 | 40 | | 6 | 55 | 864 | 33156 | $125,000.00 |
| 7 | 114 | 722 | 6398 | $33,904.56 | $35,667.66 | 41 | | 7 | 21 | 761 | 32706 | $125,000.00 |
| 8 | 133 | 723 | 10626 | $50,223.16 | $63,810.19 | 30 | | 8 | 56 | 865 | 32118 | $125,000.00 |
| 9 | 151 | 724 | 7434 | $37,903.16 | $44,252.16 | 37 | | 9 | 13 | 734 | 24585 | $104,100.00 |
| 10 | 170 | 731 | 2957 | $20,623.51 | $19,102.00 | 63 | | 10 | 11 | 732 | 23500 | $99,912.28 |

*Figure 8 Ordering by decreasing population*

## Figure 9 Trying to order by Cost

Formula bar (S2): `=RANK.EQ(I2,Cost,0)`

| Cost Rank | Seq No Cost | Row Num Cost |
|---|---|---|
| 69 | 1 | 21 |
| 57 | 2 | #N/A |
| 33 | 3 | #N/A |
| 62 | 4 | #N/A |
| 42 | 5 | #N/A |
| 40 | 6 | #N/A |
| 41 | 7 | #N/A |
| 30 | 8 | #N/A |
| 37 | 9 | 13 |
| 63 | 10 | 11 |

*Figure 9 Trying to order by Cost*

## Figure 10 Ordering with a surrogate key

Formula bar (X2): `=RANK.EQ(W2,SK,1)`

| Pop Rank | Cost Rank | Seq No Cost | Row Num Cost | | SK | SK Rank | Row Num SK | Club | Pop | Cost |
|---|---|---|---|---|---|---|---|---|---|---|
| 72 | 69 | 1 | 21 | | 6972 | 72 | 22 | 762 | 74326 | $125,000.00 |
| 57 | 57 | 2 | #N/A | | 5757 | 57 | 23 | 763 | 46784 | $125,000.00 |
| 33 | 33 | 3 | #N/A | | 3333 | 33 | 45 | 831 | 41406 | $125,000.00 |
| 62 | 62 | 4 | #N/A | | 6262 | 62 | 48 | 851 | 35984 | $125,000.00 |
| 42 | 42 | 5 | #N/A | | 4242 | 42 | 46 | 841 | 34901 | $125,000.00 |
| 40 | 40 | 6 | #N/A | | 4040 | 40 | 55 | 864 | 33156 | $125,000.00 |
| 41 | 41 | 7 | #N/A | | 4141 | 41 | 21 | 761 | 32706 | $125,000.00 |
| 55 | 55 | 20 | 15 | | 5555 | 55 | 15 | 741 | 16007 | $70,991.93 |
| 7 | 1 | 21 | 43 | | 107 | 7 | 43 | 823 | 15871 | $70,467.02 |
| 1 | 1 | 22 | 26 | | 101 | 1 | 26 | 772 | 15809 | $70,227.72 |
| 2 | 1 | 23 | 57 | | 102 | 2 | 57 | 866 | 15722 | $69,891.93 |
| 61 | 61 | 24 | 60 | | 6161 | 61 | 60 | 869 | 15382 | $68,579.65 |

*Figure 10 Ordering with a surrogate key*

If we were to use the ORDER BY COST DECREASING clause in a SELECT statement, the database management system (DBMS) would not produce an error message: it would return all the rows, but the ordering of the first 8 rows is dependent on the DBMS not guaranteed to be same with another DBMS. A good practice is to specify as many ordering criteria in the ORDER BY clause to ensure a unique ordering.

With Excel, we need to use one or more other column rankings to break the ties and create a unique surrogate sorting key. In our example, we know that the population has unique values. We could then build a surrogate key SK as follows:

```
SK = Cost Rank * A + Pop Rank
```

Here, A is a value greater than the size of the set we are sorting. Figure 10 illustrates the construction of SK (with A=100). Columns L to R and rows 9 to 20 have been hidden to show the behavior of the sorting key in rows 22 to 24.

Note that we can specify the direction of the ranks of *Pop* and *Cost* that interest us with the last parameter of the RANK.EQ function (0 for decreasing or 1 for increasing). But the sorting key is normally sorted increasingly.

Also note that if there is a possibility of having a non-unique SK, you should add more rankings in its calculation. You can stop when you add a ranking that is known to be unique. As a last resort, you can use the original sequence numbers of the data. In our example, since it is possible to have two clubs with the same population, we should use the following key:

```
SK = (Cost Rank * A + Pop Rank) * A + NoSeq
```

## VI. CONCLUSION

In this paper, we presented a set of formulas to prepare the necessary structures to select instances of variables whose values change dynamically during the spreadsheet's use. These model management formulas serve to emulate the workings of the WHERE and ORDER BY clauses of the SQL SELECT statement.

The purpose of the formulas presented in this paper is to facilitate the use of the spreadsheet as a decision support tool by avoiding manipulations by the user.

## VII. REFERENCES

[1] P. Mireault, "Structured Spreadsheet Modeling and Implementation," in *SEMS 15*, Florence, 2015: IEEE.
[2] J. Cunha, J. P. Fernandes, J. Mendes, and J. Saraiva, "MDSheet: a framework for model-driven spreadsheet engineering," in *Proceedings of the 34th International Conference on Software Engineering*, Zurich, 2012.
[3] J. Cunha, M. Erwig, J. Mendes, and J. Saraiva, "Model inference for spreadsheets," *Autom Softw Eng,* vol. 23, pp. 361-.392, 2016.
[4] P. Mireault, "Structured Spreadsheet Modelling and Implementation with Multiple Dimensions - Part 1: Modelling," presented at the EuSpRIG 2017 Conference, London, UK, 2017.
[5] P. Mireault, "Structured Spreadsheet Modelling and Implementation with Multiple Dimensions - Part 2: Implementation," presented at the EuSpRIG 2018 Conference, London, UK, 2018.
[6] P. Mireault, "Implementing GROUP BY calculations as Spreadsheet Formulas," presented at the Software Engineering Methods for Spreadsheets, Eindhoven, Netherlands, 2017.